\documentclass[prb,twocolumn,showpacs,floatfix]{revtex4}
\usepackage{amsmath,amsfonts,amssymb,graphicx,bm}

\begin{document}

\title{Exchange interaction in quantum rings and wires in the Wigner-crystal
limit}
\author{Michael M. Fogler}
\author{Eugene Pivovarov}
\affiliation{Department of Physics, University of California San Diego, La Jolla,
California 92093}
\date{\today}

\begin{abstract}

We present a controlled method for computing the exchange coupling in
correlated one-dimensional electron systems based on the relation
between the exchange constant and the pair-correlation function of
spinless electrons. This relation is valid in several independent
asymptotic regimes, including low electron density case, under the
general condition of a strong spin-charge separation. Explicit formulas
for the exchange constant are obtained for thin quantum rings and wires
with realistic Coulomb interactions by calculating the pair-correlation
function via a many-body instanton approach. A remarkably smooth
interpolation between high and low electron density results is
shown to be possible. These results are applicable to the case of
one-dimensional wires of intermediate width as well. Our method can be
easily generalized to other interaction laws, such as the inverse
distance squared one of the Calogero-Sutherland-Moser model. We
demonstrate excellent agreement with the known exact results for the
latter model and show that they are relevant for a realistic
experimental setup in which the bare Coulomb interaction is screened by
an edge of a two-dimensional electron gas.

\end{abstract}

\maketitle



\section{Introduction}
\label{sec:Introduction}

Much interest has been devoted to the spin degree of freedom in novel
one-dimensional (1D) conductors, both of linear shape (carbon
nanotubes,\cite{Saito-book} semiconductor nanowires,\cite{Huang-01}
conducting molecules\cite{Heath-03}) and recently, of a circular one
(quantum rings\cite{Lorke-00, Fuhrer-04, Bayer-03}). Parameters of
these systems, e.g., average distance between the electrons $a$, their
total number $N$, their effective mass $m$, dielectric constant
$\epsilon $, effective Bohr radius $a_{B}=\hbar ^{2}\epsilon /me^{2}$,
\textit{etc.}\/, can vary over a broad range or can be tuned
experimentally. This creates an intriguing opportunity of reaching the
Wigner-crystal (WC) limit,\cite{Schulz-93,Egger-99} $r_{s}\equiv
a/2a_{B} \gg 1$, where electrons arrange themselves into a nearly
regular lattice. According to numerical simulations,\cite{Egger-99} the
1D WC is well formed already at $r_s > 4$. The corresponding electron
densities are easily achievable but the presence of disorder has so far
hindered experimental investigations of the 1D WC
regime.\cite{Field-90} Remarkably, the unwanted disorder can apparently
be substantially suppressed in the case of carbon nanotubes
\emph{suspended\/} above a substrate. In such devices very large $r_s$
with no immediately obvious disorder effects have been recently
demonstrated.\cite{Jarillo-Herrero-04} Due to a finite length of the
nanotubes used, they contained only a few electrons ($< 30$) in the $r_s
> 4$ regime. (Such systems are commonly referred to as Wigner
molecules.) These encouraging progress on the experimental side and a
recent revival of interest to the 1D WC on the theoretical
side\cite{Matveev-04} have motivated us to undertake in the present
paper a careful investigation of the fundamental energy scales that
govern the spin dynamics in 1D Wigner crystals and molecules.

In a strict academic sense, the term ``crystal'' should not be used in
1D because the phonon-like vibrations of the putative lattice are
infrared divergent. It is nevertheless true that at $r_{s}\gg 1$ the
pair correlation function (PCF)
\begin{equation}
g(x)=\frac{1}{N n}\sum_{i\neq j}\langle \delta (x_{i}-x_{j}-x)\rangle
\label{eqn:g_intro}
\end{equation}
is sharply peaked at integer multiples of $a$. Here $x_j$ are electron
coordinates. Thus, for the study of electron correlations in Wigner
molecules or in a group of few nearby electrons in an infinite wire the
WC concept is fully adequate. Once adopted, this concept
implies\cite{Herring-62} that the low-energy dynamics of electron spins
is dominated by the nearest-neighbor Heisenberg interaction of strength
$J$ proportional to an exponentially small probability of quantum
tunneling under the Coulomb barriers that separate the adjacent
electrons. Exchanges of more distant neighbors are penalized with a much
stronger tunneling suppression and can therefore be neglected.

%
\begin{figure}
\begin{center}
\includegraphics[height=1.5in]{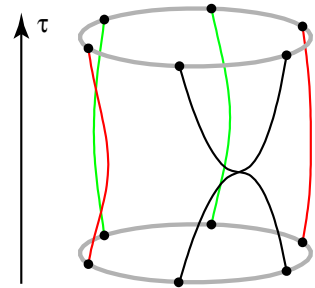}
\end{center}
\caption{(Color online) A sketch of the instanton trajectories for
the six-electron Wigner molecule on a ring.}
\label{fig:ring}
\end{figure}

As a result of the exponential smallness of the exchange coupling $J$,
the energy scales for orbital and spin dynamics are drastically
different. This \emph{strong\/} spin-charge separation is expected to
cause anomalies in many essential electron properties, e.g., ballistic
conductance\cite{Matveev-04} of quantum wires and persistent current of
quantum rings.\cite{Reimann-02} To systematically explore such
phenomena a reliable estimate of the exchange coupling $J$ is needed.
This has been an important open problem however. The smallness of $J$
makes it difficult to compute even by rather advanced computer
simulations.\cite{Reimann-02} Attempts to derive $J$ analytically (for
the nontrivial case $N>2$) have so far been based on the approximation
that neglects all degrees of freedom in the problem except the distance
between the two interchanging electrons.\cite{Hausler-96, Matveev-04} We
call this a Frozen Lattice Approximation (FLA). The accuracy of the FLA
is unclear because it is not justified by any small parameter. When a
given pair does its exchange, it sets all other electrons in motion, too
(Fig.~\ref{fig:ring}). Simple dimensional considerations (confirmed by
the calculations below) show that the maximum tunneling displacements of
the electrons nearby the first pair reach a fraction of $a$, i.e., they
are at most numerically, but not \emph{parametrically}, small. In view
of the aforementioned and other recent experimental\cite{Auslaender-05,
Claessen-02} and theoretical\cite{Cheianov-04, Fiete-04, Fiete-05} work
on spin-related effects in 1D conductors a controlled calculation of $J$
seems a timely goal. It is accomplished in this paper, where we treat
the spin exchange in a Wigner molecule (or a WC) as a true many-body
process and compute $J$ to the leading order in the large parameter
$r_s$. A brief account of this work has been reported in
Ref.~\onlinecite{Fogler-05b}. The same problem has been independently
studied by \textcite{Klironomos-05} and wherever they can be compared
our results agree.

The paper is organized as follows. The definition of the model and our
main results are given in Sec.~\ref{sec:Results}. The step-by-step
derivation for the simplest nontrivial case of a three-electron molecule
is presented in Sec.~\ref{sec:Three}. The $N > 3$ case is discussed in
Secs.~\ref{sec:Many} and \ref{sec:Instanton}. An alternative route to
the derivation of $J$ is outlined in Sec.~\ref{sec:Alternative}. The
comparison with the known analytical results for related models is done
in Sec.~\ref{sec:integrable_model}. The issues relevant to the current
experiments are addressed at the end of Sec.~\ref{sec:integrable_model}
and in Sec.~\ref{sec:Domain}. The concluding remarks are in
Sec.~\ref{section:Conclusions}.

\section{Model and main results}
\label{sec:Results}

We assume that electrons are tightly confined in the transverse
dimensions on a characteristic lengthscale (``radius'' of the wire)
$R\ll a_{B}$. In this case the energy separation $\hbar^2 / m R^2$ of
the 1D subbands greatly exceeds the characteristic Coulomb interaction
energy $e^2 / \epsilon R$. Assuming that only the lowest subband is
occupied, the Hamiltonian can be projected onto the Hilbert space of
this subband. This turns the problem into a strictly 1D one, with the
Coulomb law replaced by an effective interaction $U(r)$ that tends to a
finite value at distances $r\ll R$. The precise form of $U$ depends on
the details of the confinement. For simplicity, we adopt a
model\cite{Fogler-05a, Usukura-05}
\begin{equation}
U(r) = \frac{e^2}{\epsilon} \frac{1}{r + R},
\label{eqn:U}
\end{equation}
which nevertheless correctly captures both short- and long-range
behavior of the interaction for \emph{any\/} realistic confinement
scheme and is similar to other expressions used in the
literature,\cite{Friesen-80, Szafran-04} see more in
Sec.~\ref{sec:Domain}. Most of our calculations below pertain to the
ring geometry where $r$ stands for the chord distance, $r = (Na/\pi
)\left\vert \sin (\pi x/Na)\right\vert$, $x$ being the coordinate along
the circumference. At large $r_s$ the electrons assume a Wigner-molecule
configuration where they reside at the corners of a regular polygon. The
effective low-energy Hamiltonian of such a state is given by
\begin{equation}
H=\frac{\hbar ^{2}}{2I}L^{2}+J\sum_{j}\mathbf{S}_{j}
\mathbf{S}_{j+1}+\sum_{\alpha }n_{a}\hbar \omega _{\alpha },
\label{eqn:H}
\end{equation}
where $L$ is the center-of-mass angular momentum, $\mathbf{S}_{j}$ are
electron spins, and $n_{\alpha }$ are the occupation numbers of
``molecular vibrations.'' (The same
Hamiltonian applies to small-$N$ quantum dots.\cite{Reimann-02})
At large $r_{s}$, the total moment of inertia $I$ and the vibrational
frequencies $\omega_\alpha$ are easy to compute because they are
close to their classical values. Our task is to calculate $J$, which
is more difficult.

We show that the asymptotically exact relation exists between $J$ and
the PCF $g_0(x)$ of a \emph{spin polarized\/}
1D system. For an ultrathin wire, $\mathcal{L}\equiv \ln
({a_{B}}/{R})\gg 1$, it is particularly simple:
\begin{equation}
J = \frac{e^2 a_B^2}{2\mathcal{L}\epsilon} g_0^{\prime \prime}(0),
\quad r_{s}\gg \frac{1}{\mathcal{L}}.
\label{eqn:J_from_g_ultrathin}
\end{equation}
By virtue of Eq.~(\ref{eqn:J_from_g_ultrathin}), the calculation of $J$
reduces to an easier task of computing $g_0(x)$. For large $r_{s}$ our final
result has the form
\begin{equation}
J=\frac{\kappa }{\left( 2r_{s}\right) ^{5/4}}\frac{\pi }{\mathcal{L}}
\frac{e^{2}}{\epsilon a_{B}}\exp \left( -\eta \sqrt{2r_{s}}\,\right) ,
\quad r_{s}\gg 1,  \label{eqn:J}
\end{equation}
where the values of $\eta $ and $\kappa $ are given in
Table~\ref{tbl:Results} together with the prediction of the
FLA,\cite{Hausler-96,Matveev-04} for comparison.
[\textcite{Klironomos-05} independently obtained $\eta =2.79805(5)$ for
the case of a wire.]
\begin{table}[tbp]
\begin{ruledtabular}
\begin{tabular}{lcccccc}
$N$      &3      &4         &6         &8         &$\infty$  &$\infty$-FLA\\
$\eta$   &2.8009 &2.7988(2) &2.7979(2) &2.7978(2) &2.7978(2) &2.8168      \\
$\kappa$ &3.0448 &3.18(6)   &3.26(6)   &3.32(7)   &3.36(7)   &2.20        \\
\end{tabular}
\end{ruledtabular}
\caption{Results for Wigner molecules on a ring (finite $N$) and for wires
($N=\infty $). }
\label{tbl:Results}
\end{table}

In the remainder of the paper we give the derivation of the above results
and discuss their consequences for various experimental and theoretical
questions.


\section{Three-electron quantum ring}
\label{sec:Three}

We start with the simplest nontrivial example: three electrons on a
ring. Let $0\leq x_{j}<3a$, $j=0,1,2$ be their coordinates along the
circumference. We will compute the exchange coupling $J$ between the
$j=0$ and the $j=1$ electrons. In the FLA the third electron remains at
rest while the other two interchange, as in a classical picture.
However, we will show that the consistent calculation of $J$ requires
taking quantum fluctuations of this third electron into account.

The system has the total of three degrees of freedom: the center of mass
$x_{\text{cm}}$, the distance between the exchanging electrons $x\equiv
x_{1}-x_{0}$, and the distance between the third electron and the center
of mass, $X_{2}\equiv x_{2}-x_{\text{cm}}\ -a$. We can restrict the
variables $x$ and $X_{2}$ to the fundamental domain, $|x|<3a/2$,
$|X_{2}|<a/2$, where only the chosen two electrons can closely approach
each other. This allows us to ignore exchanges involving the $j=2$
electron, and hence, its spin. Ignoring also the irrelevant
center-of-mass motion, we obtain the Hamiltonian
\begin{equation}
H_{3} = -\frac{\hbar^2}{2\mu} \partial _{x}^{2}
      - \frac{\hbar^2}{2M} \partial_{X_2}^2
      + U_{\text{tot}}(x, X_2),
\label{eqn:H_3}
\end{equation}
where $\mu = m / 2$ and $M = 3 \mu$. The potential term
\begin{align}
U_{\text{tot}} & = U(x)+U_{\text{ext}},
\\
U_{\text{ext}} & = U\left[ (3/2)(X_{2}+a)-x/2\right]
\notag\\
& + U\left[ (3/2)(X_{2}+a)+x/2\right] ,
\notag
\end{align}
has two minima in the fundamental domain, at $x=\pm a$, $X_{2}=0$. The
minima are separated by a high potential barrier at $x=0$. They give
rise to the two lowest-energy states of the system: the spin-singlet ground
state, $\mathbf{S}_{0}+\mathbf{S}_{1}=0$, with an orbital wavefunction $\Phi
_{s}=\Phi _{s}(x,X_{2})$ and a triplet with a wavefunction $\Phi _{t}$.
Their energy splitting is the desired exchange coupling $J$. In close
analogy with Herring's classic treatment of H$_2$-molecule\cite{Herring-64}
(see also Ref.~\onlinecite{Landau-III}) for our $N=3$ ``molecule''
we obtain
\begin{equation}
J = \frac{2\hbar^2}{\mu} \int dX_{2}\left. \Phi _{1}\partial _{x}\Phi
_{1}\right\vert _{x=0},  \label{eqn:J_from_Phi_1}
\end{equation}
where the (normalized to unity) ``single-well'' wavefunction $\Phi _{1}(x,X_{2})$
is the ground-state of $H_{3}$ with a modified potential $U_{\text{tot}}\
\rightarrow U_{1}\equiv U_{\text{tot}}(\max \{x,0\},X_{2})$.
Equation~(\ref{eqn:J_from_Phi_1}) is valid\cite{Herring-64} to order $O(J^{2})$;
with the same accuracy,
\begin{equation}
\Phi _{s,t}=\frac{1}{\sqrt{2}}[\Phi _{1}(x,X_{2})\pm \Phi _{1}(-x,X_{2})].
\end{equation}

Let us discuss the form of the single-well wavefunction
$\Phi _{1}(x,X_{2})$. Near its maximum at $x=a$,
$X_{2}=0$, it is a simple Gaussian,
\begin{equation}
\Phi _{1}\left( x,X_{2}\right) \propto \exp \left[ -\frac{(x-a)^{2}}{2l^{2}}-%
\frac{M}{2\hbar }\omega (a)X_{2}^{2}\right],
\label{eqn:Phi_1_Gauss}
\end{equation}
where $l=[\hbar ^{2}/\mu U_{\text{tot}}^{\prime \prime }(a)]^{1/4}\sim
r_{s}^{-1/4}a$ is the amplitude of the zero-point motion in $x$
($U_{\text{tot}}$ written with a single argument is meant to be evaluated at
$X_{2}=0$). The quantity
\begin{equation}
\omega (x)=\left[ M^{-1}\partial _{X_{2}}^{2}U_{\text{tot}}(x)\right] ^{1/2}
\end{equation}
is real and positive in the classically forbidden region $0<x<a-l$.
Therefore, the tunneling barrier is the lowest and $\Phi _{1}$ is the
largest along the line $X_{2}=0$. Furthermore, it is easy to see that $\Phi
_{1}(x,X_{2})$ rapidly decays at $|X_{2}|\gtrsim l$. Since $l\ll a$,
the following Gaussian approximation is justified in the \emph{entire\/}
fundamental domain of $x$:
\begin{equation}
\Phi_{1}=\phi \left( x\right) \exp \left[ -\frac{M}{2\hbar }\Omega \left(
x\right) X_{2}^{2}\right].
\label{eqn:Phi_1_ansatz}
\end{equation}
Acting on this \emph{ansatz\/} with $H_{3}$ and neglecting the subleading
terms $O(l/a)$, we obtain the following equations on $\Omega(x)$ and
$\phi(x)$:
\begin{subequations}
\label{eqn:Omega-phi-3}
\begin{gather}
\partial _{x}\Omega = \frac{\Omega^2(x) - \omega^2(x)}
{\left[(2 / \mu) \Delta U_{\text{tot}}(x)\right]^{1/2}},
\label{eqn:Omega_3} \\
\left\{ \frac{\hbar^2}{2\mu} \partial _{x}^{2}-U_{\text{tot}}(x)- \frac12
\hbar \Omega(x) + E\right\} \phi (x)=0,
\label{eqn:phi_3}
\end{gather}
\end{subequations}
where $\Delta U_{\text{tot}}(x)\equiv U_{\text{tot}}(x)-U_{\text{tot}}(a)$.
We wish to comment that the basic idea of the Gaussian ansatz has been
used previously for computing exchange constants in $^{3}$\textsc{H}e
crystals,\cite{Roger-83} but the simple closed form of
Eqs.~(\ref{eqn:Omega-phi-3}) has not been demonstrated.

Comparing Eqs.~(\ref{eqn:Phi_1_Gauss}) and (\ref{eqn:Phi_1_ansatz})
we see that near $x = a$ function $\phi(x)$ takes the form,
\begin{equation}
\phi \left( x\right) =\left[ \frac{M\Omega \left( a\right) }{\pi ^{3/2}\hbar
l}\right] ^{1/2}e^{-\left( x-a\right) ^{2}/2l^{2}},
\end{equation}
while $\Omega(x)$ satisfies the boundary condition $\Omega(a) = \omega
(a)$. Resolving the $0/0$ ambiguity in Eq.~(\ref{eqn:Omega_3}) by
L'H\^opital's rule we furthermore obtain
\begin{equation}
\Omega ^{\prime }\left( a\right) =\frac{\omega ^{\prime }\left( a\right) }
{1 + \left\{ \left( \frac{M}{16 \mu }\right) \left[ 1+U^{\prime \prime }\left(
a\right) /U_{\text{ext}}^{\prime \prime }\left( a\right) \right] \right\}
^{1/2}}.
\end{equation}
Finally, using this formula and straightforward algebraic manipulations one
can fix the constant term $E$ in Eq.~(\ref{eqn:phi_3}) to be
\begin{equation}
E = U_{\text{tot}}(a)+ \frac{\hbar}{2} \left[ \omega (a)+\omega _{0}\right],
\quad \omega_0 \equiv \frac{\hbar}{\mu l^2}.
\end{equation}
%
%
%
Next consider the interior of the classically forbidden region, $x \ll a$.
Since the tunneling barrier $\Delta U_{\text{tot}}(x)$ is large here,
$\Omega$ is a slow function of $x$. Taking advantage of the following
expression for the PCF of a spin-polarized molecule,
\begin{equation}
g_0(x)\equiv 2\!\int \prod_{j=2}^{N-1}dX_{j}\Phi _{t}^{2}\left( x,X_{2},\ldots
,X_{N-1}\right),
\quad |x|<\frac{3a}{2},
\label{eqn:g_def}
\end{equation}
we find that Eq.~(\ref{eqn:J_from_Phi_1}) entails the relation
\begin{equation}
J = \frac{\hbar^2}{4\mu} \frac{\phi(0)}{\phi^\prime(0)}
    g_0^{\prime \prime}(0).
\label{eqn:J_from_g_triplet}
\end{equation}
Anticipating the discussion in Sec.~\ref{sec:Many},
Eq.~(\ref{eqn:g_def}) is written for an arbitrary $N>2$, with the
notation $X_{j}\equiv x_{j}-x_{\text{cm}}\ +(N-1-2j)(a/2)$ being used;
the PCF is normalized as appropriate in the WC limit,
\begin{equation}
\int_{0}^{3a/2} g_0(x) dx = 1.
\label{eqn:g_normalization}
\end{equation}
%

%
\begin{figure}
\begin{center}
\includegraphics[height=1.3in]{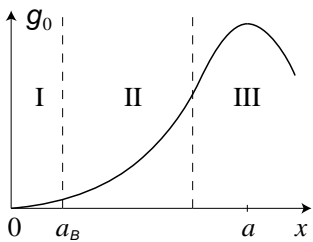}
\end{center}
\caption{The PCF of a spin-polarized system (schematically). Regions I, II,
and III are described in the main text.}
\label{fig:PCF}
\end{figure}

The dependence of $g_0$ on $x$ that results from this formalism is
sketched in Fig.~\ref{fig:PCF}. Near the $x = a$ maximum (region III),
$g_0(x)$ is given by [cf.~Eq.~(\ref{eqn:Phi_1_Gauss})]
\begin{equation}
g_0(x)=\left( 1/\sqrt{\pi }\,l\right) \exp \left[ -(x-a)^{2}/l^{2}\right] .
\label{eqn:g_Gauss}
\end{equation}
In the region II the quasiclassical approximation applies. In particular, at
$a_{B}\ll x\ll a$ the result for $g_0(x)$ can be written in terms of the
tunneling action,
\begin{equation}
S_{3}(x)=\frac{1}{\hbar }\int\limits_{x}^{a}dy\left[ 2\mu \Delta U_{\text{tot%
}}(y)\right] ^{1/2},  \label{eqn:S_3}
\end{equation}
and the appropriate prefactor, as follows:
\begin{align}
g_0(x)& =\frac{a}{l^{2}}\left[ \frac{1}{2\pi }\frac{\Omega (a)}{\Omega (x)}%
\frac{\hbar \omega _{0}}{U(x)}\right] ^{1/2}e^{\xi (x)-2S_{3}(x)},
\label{eqn:g_II} \\
\xi (x)& =\int\limits_{x}^{a}dy\left\{ \frac{\omega _{0}+\Omega (a)-\Omega
(y)}{\left[ (2/\mu )\Delta U_{\text{tot}}(y)\right] ^{1/2}}-\frac{1}{a-y}%
\right\} .  \label{eqn:xi}
\end{align}
In comparison, the FLA\cite{Matveev-04} amounts to replacing
$\Omega(x)$ by a constant in the above equations, thereby effectively
ignoring the quantum fluctuations of the dynamical variable $X_2$.
Finally, in an ultrathin wire, $\mathcal{L} \gg 1$, there is also region
I, $x\lesssim a_{B}$, where the quasiclassical approximation breaks
down. Fortunately, at such $x$, Eq.~(\ref{eqn:phi_3}) can be simplified,
as $\Omega (x)\simeq \Omega (0)$ and $U_{\text{tot}} (x)\simeq
U(x)+2U(3a/2)$. This enables us to express $\phi (x)$ and $g_0(x)$ in
terms of the Whittaker functions,\cite{Gradshteyn-Ryzhik} similar to
Ref.~\onlinecite{Fogler-05a} (see also Appendix~\ref{sec:U_sigma}).
Using such a representation, it is easy to show that
\begin{equation}
\phi (0)/\phi ^{\prime }(0)\simeq a_{B}/\mathcal{L},\quad r_{s}\gg 1,
\label{eqn:phi_log_derivative}
\end{equation}
which, combined with Eq.~(\ref{eqn:J_from_g_triplet}), yields
Eq.~(\ref{eqn:J_from_g_ultrathin}). With a bit
more algebra, one can match Eq.~(\ref{eqn:g_II}) with another formula in
region I, leading finally to Eq.~(\ref{eqn:J}) with $\eta $ and $\kappa $
given by
\begin{align}
\eta & =\frac{2S_{3}(0)}{\sqrt{2r_{s}}}=2\int\limits_{0}^{a}\frac{dx}{a}%
\left[ \frac{\epsilon a}{e^{2}}\,\Delta U_{\text{tot}}(x)\right] ^{1/2},
\label{eqn:eta_3} \\
\kappa & =\frac{2^{5/4}}{\sqrt{\pi }}\,e^{\xi (0)}\sqrt{\frac{\Omega (a)}
{\Omega (0)}}\left[ \frac{\epsilon a^{3}}{e^{2}}\,U_{\text{tot}}^{\prime
\prime }(a)\right] ^{3/4}.  \label{eqn:kappa_3}
\end{align}
Thus, for the $N=3$ case we were able to reduce the original complicated
three-body eigenvalue problem to routine operations of solving an
\emph{ordinary\/} differential equation~(\ref{eqn:Omega_3}) and taking two
quadratures, Eqs.~(\ref{eqn:xi}) and (\ref{eqn:eta_3}). The resultant $\eta$
and $\kappa$ are listed in Table~\ref{tbl:Results}. In comparison, the FLA%
\cite{Matveev-04} underestimates $\kappa $ by about $50\%$. It gets $\eta$
correctly but only for $N=3$, see more below.

One more important comment is in order. The antisymmetry of the total
fermion wavefunction imposes certain selection rules\cite{Maksym-96} for the
allowed values of $L$ [see Eq.~(\ref{eqn:H})] at a given total spin $S$.
Thus, the lowest-energy $L$ eigenstates for the two possible spin states of
the $N = 3$ system, $S = 1/2$ and $S = 3/2$, are, respectively, $|L| = 1$
and $0$. Since $J \ll \hbar^2 / I$ at large $r_s$, the ground-state of the
system is the $L = 0$ spin-quartet, in agreement with prior numerical work.%
\cite{Reimann-02, Usukura-05}


\section{Case $N>3$}
\label{sec:Many}

In a system of more than three electrons, the Hamiltonian that governs the
important degrees of freedom, $x$ and $\mathbf{X}=(X_{2},\ldots
,X_{N-1})^{\dagger }$, becomes
\begin{equation}
H_{N}=-\frac{\hbar ^{2}}{2\mu }\partial _{x}^{2}-\frac{\hbar ^{2}}{2}\left(
\mathbf{M}^{-1/2}\partial _{\mathbf{X}}\right) ^{\dagger }\left( \mathbf{M}%
^{-1/2}\partial _{\mathbf{X}}\right) +U_{\text{tot}},  \label{eqn:H_N}
\end{equation}
where

\begin{equation}
M_{ij}^{-1/2}=\frac{1}{\sqrt{m}}\frac{\delta _{ij}-\left( 1-\sqrt{2/N}%
\right) }{N-2}.
\end{equation}
Again, the potential energy has two minima separated by a large barrier.
The single-well function $\Phi _{1}(x,\mathbf{X})$ can be sought in the form
\begin{equation}
\Phi _{1}=\phi (x)\exp \left[ -\frac{1}{2\hbar }\Delta \mathbf{X}^{\dagger }%
\mathbf{M}^{1/2}\bm{\Omega}(x)\mathbf{M}^{1/2}\Delta \mathbf{X}\right] ,
\label{eqn:Phi_1}
\end{equation}
where $\Delta \mathbf{X}=\mathbf{X}-\mathbf{X}^{\ast }$. In the language of
the quantum tunneling theory, $\mathbf{X}^{\ast }(x)$ is the instanton
trajectory and $\bm{\Omega}(x)$ is a matrix that controls Gaussian
fluctuations around the instanton. Our goal is to compute them. The usual
route is to parametrize the dependence of $\mathbf{X}^{\ast }$ on $x$ in
terms of an ``imaginary-time'' $\tau $, in
which case $x(\tau )$ and $\mathbf{X}^{\ast }(\tau )$ must minimize the
action functional
\begin{equation}
S_{N}=\int\limits_{0}^{\infty }\frac{d\tau }{\hbar }\left[ \frac{\mu }{2}%
(\partial _{\tau }x)^{2}+\frac{1}{2}(\partial _{\tau }\mathbf{X})^{\dagger }%
\mathbf{M}\,\partial _{\tau }\mathbf{X}+\Delta U_{\text{tot}}\ \right]
\label{eqn:S_N}
\end{equation}
subject to the boundary conditions $x(0)=0$, $x(\infty )=a$, and $\mathbf{X}
(\infty )=0$. Henceforth $U_{\text{tot}}$ is always meant to be evaluated on
the instanton trajectory and $\Delta U_{\text{tot}}$ stands for the
difference of its values at a given $\tau $ and at $\tau =\infty $.
Repeating the steps of the derivation for the $N=3$ case, we find that
Eq.~(\ref{eqn:Omega_3}) for $\bm{\Omega}$ is still valid once we define
$\bm{\omega}$ to be a positive-definite matrix such that $\bm{\omega}^{2}=
\mathbf{M}^{-1/2}\bm{\Xi}\mathbf{M}^{-1/2}$, where $\bm{\Xi}$ is the matrix
of the second derivatives $\Xi _{ij}=\partial _{X_{i}}\partial _{X_{j}}
U_{\text{tot}}$. This equation has an equivalent but more elegant form in terms
of $\tau $:
\begin{equation}
\partial _{\tau }\bm{\Omega}=\bm{\Omega}^{2}(\tau )-\bm{\omega}^{2}(\tau ).
\label{eqn:Omega}
\end{equation}
Also, for practical calculations, it is convenient to take advantage of a
formula
\begin{equation}
\text{tr}\,\left( \mathbf{M}^{-1}\bm{\Xi}\right) +\frac{1}{\mu }
\frac{\partial ^{2}}{\partial x^{2}}U_{\text{tot}}\left( x\right) =
\frac{1}{2\mu }\sum\limits_{\substack{ i,j=0  \\ i\neq j}}^{N-1}
U^{\prime \prime }\left( x_{i}-x_{j}\right) .
\end{equation}

As for Eqs.~(\ref{eqn:phi_3}) and (\ref{eqn:xi}), they require only the
replacement $\Omega \rightarrow \text{tr}\,\bm{\Omega}$,
\begin{equation}
\left\{ \frac{\hbar ^{2}}{2\mu }\partial _{x}^{2}-U_{\text{tot}}(x) -
\frac{\hbar }{2}\,\text{tr}\,\bm{\Omega}(x) + E\right\} \phi (x)=0,
\end{equation}
while in Eq.~(\ref{eqn:kappa_3}) one has to replace $\Omega $ with $\det
\bm{\Omega}$.

Finally, instead of Eq.~(\ref{eqn:eta_3}) we have $\eta =2S_{N}/\sqrt{2r_{s}}
$, consistent with the notion that the main exponential dependence of $J$ is
always determined by the tunneling action.


\section{Calculation of the instanton}
\label{sec:Instanton}

A few properties of the instanton follow from general considerations. The
dimensional analysis of Eq.~(\ref{eqn:S_N}) yields $S_{N}\propto \sqrt{r_{s}}
$, so that $\eta $ is indeed just a constant. Also, from the symmetry of the
problem, $X_{N+1-j}(\tau )=-X_{j}(\tau )$. Thus, in the special case of
$N=3$, the instanton trajectory is trivial: $X_{2}\equiv 0$, i.e., the $j=2$
electron does not move. This is why we were able to compute $S_{3}$ in a
closed form, Eq.~(\ref{eqn:S_3}). For $N>3$ the situation is quite
different: all electrons [except $j=(N+1)/2$ for odd $N$] do move. In order
to investigate how important the motion of electrons distant from the
$j=0,1$-pair is let us consider the $N=\infty $ (\emph{quantum wire\/}) case,
where the far-field effects are the largest. If $X_{j}$'s were small, we
could expand $\Delta U_{\text{tot}}$ in Eq.~(\ref{eqn:S_N}) to the second order
in $X_{j}$ to obtain the harmonic action
\begin{equation}
S_{h}=\frac{1}{2}\frac{m}{\hbar }\int \frac{dk}{2\pi }\int \frac{d\omega }
{2\pi }\left\vert u_{k\omega }\right\vert ^{2}\left[ \omega ^{2}+\omega
_{p}^{2}(k)\right] ,  \label{eqn:action_harmonic}
\end{equation}
where $u_{k\omega }$ is the Fourier transform of $u_{j}(\tau )\equiv
x_{j}-x_{j}^{0}$, electron displacement from the classical equilibrium
position $x_{j}^{0}\equiv (j-1/2)a$, $j\in \mathbb{Z}$, and
\begin{equation}
\omega_{p}(k)\simeq s_{0} k \ln^{1/2}\left(\frac{4.15}{ka}\right),
\quad s_{0} \equiv \sqrt{\frac{e^2}{\epsilon \mu a}}
\end{equation}
is the plasmon dispersion in the 1D WC (the logarithmic term is due to
the long-range nature of the Coulomb interaction). Minimization of
$S_{h}$ with the specified boundary conditions yields
\begin{equation}
u_{j}(\tau ) \propto \frac{v x_j^0}{(x_j^0)^2 + v^2 \tau^2},\quad
v \simeq \frac{s_0}{2}
          \ln \frac{(x_{j}^{0})^2 + s_0^2 \tau^2}{a^2}.
\end{equation}
Substituting this formula into Eq.~(\ref{eqn:action_harmonic}), we
find that the contributions of distant electrons to $S_{h}$ rapidly
decay with $|j|$. Since $u_{j}(\tau )$'s are small at large $j$ and
$\tau $ we expect that these $j$- and $\tau $-dependencies are rendered
correctly by the harmonic approximation. Thus, a fast convergence of
$\eta $ to its thermodynamic limit is expected as $N$ increases.

\begin{figure}
\begin{center}
\includegraphics[height=1.75in]{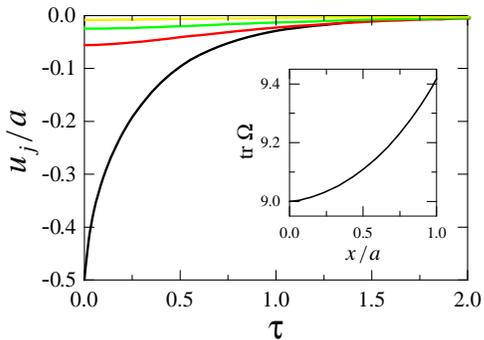}
\end{center}
\caption{(Color online) The instanton trajectories of $1\leq j\leq 4$
electrons in the $N = 8$ Wigner molecule on a ring. Inset: $\text{tr}\,
\bm{\Omega}(x)$. The units of $\protect\tau$ and $\bm{\Omega}$ are $\protect%
\sqrt{2} a / s_0$ and its inverse.}
\label{fig:Instanton}
\end{figure}

Encouraged by this conclusion, we undertook a direct numerical minimization
of $S$ for the set of $N$ listed in Table~\ref{tbl:Results} using standard
algorithms of a popular software package \textsc{MATLAB}. The optimal
trajectories that we found for the case $N = 8$ are shown in
Fig.~\ref{fig:Instanton}. In agreement with our earlier statement, $u_j(0)$
reach a finite fraction of $a$. This collective electron motion lowers the
effective tunneling barrier and causes $\eta$ to drop below its FLA value,
although only by $0.7\%$, see Table~\ref{tbl:Results}. Further decrease of
$\eta$ as $N$ increases past $8$ is beyond the accuracy of the employed
minimization procedure.

Let us now discuss the prefactor $\kappa$. In the inset of
Fig.~\ref{fig:Instanton} we plot $\text{tr}\,\bm{\Omega}(x)$ computed by
solving Eq.~(\ref{eqn:Omega}) numerically. To reduce the calculational
burden, we set $\mathbf{X}^\ast(\tau) \to 0$ instead of using the true
instanton trajectory. The error in $\kappa$ incurred thereby is $\sim
2\%$ (see Sec.~\ref{sec:integrable_model}). In comparison, the FLA, where
$\text{tr}\,\bm{\Omega}(x) = \text{const}$, yields $\kappa$ about $50\%$
smaller than the correct result, similar to $N = 3$.

%


\section{Alternative derivation of the exchange constant}
\label{sec:Alternative}

The relation~(\ref{eqn:J_from_g_ultrathin}) between the exchange
constant $J$ and the PCF of the spinless system --- which is one of our
main results --- was derived in Sec.~\ref{sec:Three} assuming $r_s \gg
1$. In this section we show that in ultrathin wires, $\mathcal{L} \equiv
\ln (a_B / R) \gg 1$, this relation holds under a more general
condition: with an accuracy $O(1 / \mathcal{L})$, it remains valid as
long as the \emph{strong\/} spin-charge separation persists, $J \ll E_F
\sim \hbar^2 / m a^2$. In particular, it applies at $r_s \sim 1$, i.e.,
in a parametric regime which has been notoriously difficult for a
controlled theoretical analysis. Therefore, we think that
Eq.~(\ref{eqn:J_from_g_ultrathin}) amounts to some definite progress.
The fundamental reason for the broad domain of validity of
Eq.~(\ref{eqn:J_from_g_ultrathin}) is that in 1D the $1/r$ fall-off of
the Coulomb potential~(\ref{eqn:U}) constitutes a sufficiently rapid
decay. In fact, the relations of type of
Eq.~(\ref{eqn:J_from_g_ultrathin}) are generic for 1D models in which
electrons interact via the potential $U(r) \sim 1 / r^\alpha$ with
$\alpha \geq 1$. Known to us physical realizations of $\alpha > 1$ cases
include (i) $\alpha = 3$, which occurs when the bare Coulomb interaction
is screened by a nearby metallic plane,\cite{Fogler-05a} and (ii)
$\alpha = 2$, where the screening is accomplished by a half-plane, see
Sec.~\ref{sec:integrable_model} below.

In order to derive Eq.~(\ref{eqn:J_from_g_ultrathin}) for the unscreened
Coulomb interaction ($\alpha = 1$) we make a connection with a previous
work of one of us\cite{Fogler-05a} where it was shown that in ultrathin
wires an unusual correlated regime --- Coulomb Tonks gas (CTG) ---
exists in the window $1 / \mathcal{L} \ll r_s \ll 1$. In the CTG, unlike
in a WC, the long-range tails of the Coulomb potential have no important
effect on the spin degree of freedom. To the leading order in $r_s$, the
Coulomb interaction~(\ref{eqn:U}) acts simply as a strong short-range
repulsion. This allows one to characterize its effect solely in terms of
the transmission coefficient $t(q)$ for a two-electron collision with
the relative momentum $q$. Based on this observation, the following
formula for $J$ was derived\cite{Fogler-05a}
\begin{equation}
J = \frac{\hbar^2}{m} \int \frac{d q}{2 \pi} \tilde{g}_0(q) q\,
    \text{Im}\,t(q),\quad J \ll E_F.
\label{eqn:J_CTG_from_t}
\end{equation}
To see how it leads to Eq.~(\ref{eqn:J_from_g_ultrathin}) we
first note that for the potential~(\ref{eqn:U})
the transmission coefficient has the form\cite{Fogler-05a}
\begin{equation}
t(q) = \frac{i q}{i q - (m / \hbar^2) c(q)},
 \quad q \gg \frac{1}{a_B},
\label{eqn:t_q}
\end{equation}
where $c(q) = 2 (e^2 / \epsilon) \ln \left(1 / 2 q R\right)$. The slow
(in this case, logarithmic) dependence of $c$ on $q$ is crucial here and
it is precisely due to the aforementioned fast decay of $U(r)$ with
distance. As explained in Ref.~\onlinecite{Fogler-05a}, in the CTG
regime the integral in Eq.~(\ref{eqn:J_CTG_from_t}) is dominated by $q
\sim 1 / a$; therefore, to the order $O(1 / \mathcal{L})$ one can
replace $c(q)$ by $c(1 / a)$ and then by $2 (e^2 / \epsilon)
\mathcal{L}$. Finally, it is permissible to neglect $i q$ in the
denominator of Eq.~(\ref{eqn:t_q}), thereupon
Eq.~(\ref{eqn:J_CTG_from_t}) becomes identical to
Eq.~(\ref{eqn:J_from_g_ultrathin}).

Our next task is to show that Eq.~(\ref{eqn:J_from_g_ultrathin}) holds
not only in the two limiting case (CTG and WC) but also everywhere in
between. We do so with the help of a formalism that can be considered a
continuum analog of the Schrieffer-Wolfe transformation in the Hubbard
model, in particular, in its 1D version.\cite{Ogata-90}

Let $j\in \mathbb{Z}$ label electrons in the order of increasing
instantaneous coordinates $x_{j}$. Then, by construction, the coupling
constant $J$ refers to the exchanges of nearest-neighbors, say, $j = 0$
and $j = 1$. This coupling is suppressed because in order to exchange,
the two electrons have to approach each other very closely and at that
moment their mutual Coulomb repulsion $U(x)$ is very strong. Let $a_c$
be a suitable short-range cutoff. Since we are interested
in the physics at the energy scale $J$, it should be possible in
principle to ``integrate out'' the processes at the much higher energy
scale $U(a_c)$. In doing so one would trade the bare Hamiltonian $H$ for
a renormalized one, $H + H_\sigma$, which however yields the same value
of the effective coupling $J$. Traditionally, this renormalization
program is accomplished by incrementally increasing $a_c$ from zero to
$a$, the natural upper limit. In particular, the
Hamiltonian~(\ref{eqn:H}) can be viewed as the final result of such a
renormalization. However, it is fully legitimate to stop the process
while $a_c$ is still much smaller than $a$. At this stage the charge
correlations at scales shorter than $a_c$ can not longer be studied
because the corresponding degrees of freedom are already integrated out.
This implies that the form of $H_\sigma$ is not unique: $H_\sigma$ may
differ in their structure on scales $x < a_c$ as long as the resultant
$J$ is unchanged. Therefore, it should be possible to choose $H_\sigma$
in the form
\begin{equation}
H_{\sigma} = U_{\sigma }(x)(\mathbf{S}_{0}\mathbf{S}_{1}-1/4)
           + \Delta c \delta(x).
\label{eqn:H_sigma}
\end{equation}
It is especially convenient to take the limit $\Delta c \rightarrow
\infty$, in which electrons behave as impenetrable particles. Although
thus renormalized Hamiltonian forbids the pair-exchange in the real
space, if the choose $U_\sigma(x)$ correctly, the spin exchange rate $J$,
which is the only observable manifestation of an interchange of
identical particles, would be preserved. A familiar example of such a
$H_\sigma$ on a lattice is the added term
\begin{equation}
\frac{4 t^2}{U} \left(\mathbf{S}_{i}\mathbf{S}_{i+1}
 - \frac14 n_{i}n_{i+1}\right)
\label{eqn:H_sigma_Hubbard}
\end{equation}
complemented with the no-double-occupancy constraint (i.e., $U \to U +
\Delta U = \infty$) in the theory of a large-$U$ 1D Hubbard
model.\cite{Ogata-90}

The choice of $U_\sigma(x)$ in Eq.~(\ref{eqn:H_sigma}) is again not
unique. To readily get the result we want, we require
that (i) $U_{\sigma }(x)$ decays exponentially with $x$ and (ii)
its effect on any wavefunction $\Phi(x)$ that vanishes at $x = 0$ is
adequately described by the first-order perturbation theory. Under
such conditions the spin dynamics is captured
correctly to the leading order in $1 / \mathcal{L}$ as long as
$U_\sigma(x)$ satisfies the constraint
\begin{equation}
I_2 \equiv \int\limits_{0}^{\infty} d x\, x^2 U_\sigma(x)
    = \frac{\hbar^2}{m} \frac{a_{B}}{\mathcal{L}}.
\label{eqn:U_sigma}
\end{equation}
The proof is relegated to Appendix~\ref{sec:U_sigma}. From this point
the derivation of Eq.~(\ref{eqn:J_from_g_ultrathin}) is straightforward.
After the renormalization, the infinitely strong $\delta$-function
included in $H_\sigma$ causes the orbital wavefunction of the
system to vanish at $x = 0$, so that the perturbation theory in
$U_\sigma$ now applies. To the leading order we can compute the
sought exchange constant $J$ by averaging $U_{\sigma}$ over
such a wavefunction or equivalently over the renormalized PCF
$g_{\text{ren}}(x)$:
\begin{equation}
J = \frac12 \int_{-\infty}^{\infty }U_{\sigma }(x)g_{\text{ren}}(x)dx.
\label{eqn:J_from_g_ren}
\end{equation}
(The factor $1/2$ accounts for the two-body nature of the interaction.)
Let us discuss $g_{\text{ren}}(x)$ in more detail. First of all, under the
conditions of a strong spin-charge separation the PCF a spinful ($g$)
and spinless ($g_0$) systems are always very similar. Before the
renormalization (for the original Hamiltonian) both functions are peaked
at $x \sim a$ and rapidly decrease as $x \to 0$, albeit only $g_0(x)$
vanishes at $x = 0$, while $g(0)$ is simply very small. Once $H_\sigma$
is added, $g(x)$ gets replaced by the function $g_{\text{ren}}(x)$, which
vanishes at $x = 0$ and thus is nearly identical to the original $g_0$
at \emph{all\/} $x$. The PCF $g_0(x)$ of the spinless system does not
change appreciably. Since $U_{\sigma }(x)$ is short-range, the integral
in Eq.~(\ref{eqn:J_from_g_ren}) is dominated by small $x$. Therefore, we
can substitute the expansion $g_{\text{ren}}(x) \simeq g_{0}(x) \simeq
g_{0}^{\prime \prime }(0)x^{2}/2$ in place of the full $g_{\text{ren}}(x)$.
Comparing the result with Eq.~(\ref{eqn:U_sigma}), we arrive at
Eq.~(\ref{eqn:J_from_g_ultrathin}).

This alternative route to the derivation of
Eq.~(\ref{eqn:J_from_g_ultrathin}) has the advantage of being valid in a
broader range of $r_s$ compared to Herring's method\cite{Herring-64}
originally designed for the quasiclassical case $r_s \gg 1$. It may also
be more intuitive because the Hubbard-model
expression~(\ref{eqn:H_sigma_Hubbard}) is widely known. Finally, it can
be easily generalized to other interaction types, as discussed in the
next section.


\section{An integrable model with a screened Coulomb interaction}
\label{sec:integrable_model}

Our instanton calculation in Sec.~\ref{sec:Many} is expected to be
asymptotically exact in the limit of large $r_s$. However it is a good
practice to verify whenever possible that any given calculation is free
of inadvertent arithmetic or coding mistakes. With this in mind we
applied our method also to the interaction law
\begin{equation}
U(r) = \frac{\hbar^2}{m} \frac{\lambda (\lambda -1)}{(r + R)^2},
\label{eqn:U_CSM}
\end{equation}
which is similar to the celebrated Calogero-Sutherland-Moser (CSM)
model,\cite{Sutherland-71} except for the short-range cutoff $R$ in
$U(r)$. Without this cutoff (i.e., for $R = 0$) the potential $U$ is
impenetrable, so that $J = 0$. With a finite cutoff a weak tunneling
through the potential barrier $U(r)$ is possible, which imparts the
system with a spin dynamics and has interesting experimental
implications (see Appendix~\ref{sec:CSM_law}). But to finish with
the theoretical part,
let us first focus on the original CSM model, $R = 0$. This is a good
test case because, on the one hand, a number of exact analytical results
are available here. On the other hand, the model also possesses a WC
regime, at $\lambda \gg 1$, so that the corresponding PCF $g_0(x)$
should be calculable by our method. According to the exact
results,\cite{Sutherland-71, Forrester-92} the PCF behaves as
\begin{equation}
g_0(x)\simeq \frac{\varkappa \sqrt{\lambda}}{a}
\left(C \frac{x}{a}\right)^{2\lambda },
\quad x\rightarrow 0.
\label{eqn:g_CSM_exact}
\end{equation}
For the three-electron molecule on a ring the coefficients $C$ and
$\varkappa$ can be computed in a straightforward manner directly
from the exact\cite{Sutherland-71} three-body wavefunction:
\begin{equation}
N = 3,\:\: \lambda \gg 1:\quad C = \frac{8\pi}{9 \sqrt{3}},\quad
                    \varkappa = \sqrt{\frac{8\pi}{27}}.
\label{eqn:C_3}
\end{equation}
For $N = \infty$ the derivation is much more
involved, but the result\cite{Forrester-92, Ha-94} is also known:
\begin{equation}
\frac{C}{2 \pi \lambda} =  \left[
\frac{\Gamma^3(\lambda + 1) \sqrt{{3} / {\pi \lambda}}}
{\Gamma(2 \lambda + 1) \Gamma(3 \lambda + 1)}\right]^{1 / (2 \lambda)},
\quad \varkappa =\sqrt{\frac{\pi}{3}},
\label{eqn:C}
\end{equation}
where $\Gamma(z)$ is Euler gamma-function.\cite{Gradshteyn-Ryzhik}
In the WC limit the first formula reduces to
\begin{equation}
C = \frac{e\pi}{3 \sqrt{3}},\quad \lambda \gg 1.
\label{eqn:C_infty}
\end{equation}
The calculation of $g_0(x)$ by the instanton method
is virtually the same as in Sec.~\ref{sec:Instanton} except for two
minor changes. First, the quasiclassical approximation is valid down to
$x=0$, i.e., there is no region I in Fig.~\ref{fig:PCF}. Second, to
handle the logarithmic divergence of the tunneling action, inherent to
the CSM model, one has to impose a different boundary condition for the
instanton, $x(\tau =0)=z$, which makes $S_{N}$ a function of $z$
[similar to $S_{3}$ being a function of $x$ in Eq.~(\ref{eqn:S_3})]. At
small $z$ it behaves as
\begin{equation}
S_{N}\simeq \sqrt{\lambda (\lambda -1)}\,[\ln (a/z)-\Delta S].
\label{eqn:Delta_S}
\end{equation}
Replacing $\sqrt{\lambda (\lambda -1)}$ by $\lambda -1/2$, we find [cf.
Eqs.~(\ref{eqn:eta_3}) and (\ref{eqn:kappa_3})]
\begin{equation}
C=e^{\Delta S},\quad \varkappa =\frac{e^{\xi (0)-\Delta S}}{\sqrt{\pi }%
\left( l\sqrt{\lambda }/a\right) ^{3}}\sqrt{\frac{\det \mathbf{\Omega }(a)}
{\det \mathbf{\Omega }(0)}}  \label{eqn:C_varkappa_CSM}
\end{equation}
(it is easy to see that $\varkappa $ is a $\lambda $-independent constant).

Remarkably, for $N=3$ the solution of Eq.~(\ref{eqn:Omega_3}) and all the
necessary quadratures can be done analytically. Using the expression for
total potential energy
\begin{align}
U_{\text{tot}}\left( x\right) &=
\frac{\pi^2}{9} \frac{\hbar^2}{m} \frac{\lambda (\lambda - 1)}{a^2}
\notag\\
&\times \left[ \frac{1}{\sin^2 \left( \pi x/3a\right)}
 + \frac{2}{\cos^2 \left( \pi x/6a\right)} - 4\right],
\end{align}
we derive (for $\lambda \gg 1$)
\begin{equation}
\begin{split}
\phi \left( x\right) & =\left[ 4\sin \left( \frac{\pi x}{3a}\right) \left(
1+\cos \frac{\pi x}{3a}\right) \right] ^{\lambda }, \\
\Omega \left( x\right) & = \frac{\pi^2}{3} \frac{\hbar }{m a^2}
\frac{\lambda }{\cos^2 \left( \pi x/6a\right)},
\end{split}
\end{equation}
which allows one to recover the exact result Eq.~(\ref{eqn:C_3}).

To check if we can also reproduce Eq.~(\ref{eqn:C_infty}) for $N =
\infty$, we calculated $\Delta S$ and $\varkappa $ numerically for
$4\leq N\leq 10$ and fitted them to cubic polynomials in $1/N$.
Extrapolating the fits to $N=\infty $, we obtained $\exp (\Delta
S)=1.6438(4)$ and $\varkappa =1.04$ compared to the exact values
$1.6434$ and $1.02$, respectively. We attribute the $2\%$ discrepancy in
$\varkappa $ to our choice not to use the true instanton trajectory in
Eq.~(\ref{eqn:Omega}). Apparently, our method has successfully passed
this test; thus, our results for the unscreened Coulomb interaction
(Table~\ref{tbl:Results}) should also be reliable.

Let us now consider the case with a small but non-zero cutoff, $0 < R
\ll a$. As explained in Appendix~\ref{sec:CSM_law}, this model can be
relevant for the existing experimental setup of \textcite{Auslaender-05}
provided the electron density can be made
low enough. Therefore, the calculation of $J$ has an independent
significance. This calculation is again similar to that for the case
of the bare Coulomb interaction, Sec.~\ref{sec:Alternative}, except that
instead of Eq.~(\ref{eqn:U_sigma}) the constraint on the effective
potential becomes
\begin{equation}
 I_{2 \lambda} \equiv \int_{0}^{\infty }dx\, x^{2\lambda}
 U_{\sigma }\left( x\right) = \frac{\pi}{\lambda -1} \frac{\hbar^2}{m}
 R^{2\lambda - 1}.
\label{eqn:U_sigma_CSM}
\end{equation}
(For derivation, see Appendix~\ref{sec:U_sigma}.) Consequently,
Eq.~(\ref{eqn:J_from_g_ultrathin}) is replaced with
\begin{equation}
J = I_{2 \lambda} a_c^{-2\lambda} g_0(a_c),\quad R \ll a_c \ll a.
\label{eqn:J_from_g_CSM}
\end{equation}
Since $R$ is parametrically smaller than $a$, it is permissible to
substitute Eqs.~(\ref{eqn:g_CSM_exact}) and (\ref{eqn:C}) into the
last formula, which yields
\begin{align}
J &= \gamma \frac{\hbar^2}{m a^2}
     \left(\frac{R}{a}\right)^{2\lambda - 1},
\label{eqn:J_CSM}\\
\gamma &= \frac{1}{2 \lambda (\lambda - 1)}
       \frac{(2 \pi \lambda)^{2\lambda + 1} \Gamma^3(\lambda + 1)}
            {\Gamma(2 \lambda + 1) \Gamma(3 \lambda + 1)}.
\label{eqn:gamma}
\end{align}
From Eq.~(\ref{eqn:J_CSM}) one concludes that the strong spin-charge
separation in the model in hand may arise for an arbitrary $\lambda > 1$
provided ${R} / {a} \ll \min\{1,\,\lambda - 1\}$. In the experimental
setup of \textcite{Auslaender-05}, where this
model can be realized physically, we estimate $R \approx 0.7 a_B$ and
$\lambda = 1.5$--$2$, see Appendix~\ref{sec:CSM_law}, so that $\gamma =
36$--$40$. Our formula predicts that as the electron density is reduced,
a fairly rapid \emph{algebraic\/} fall-off Eq.~(\ref{eqn:J_CSM}) of $J$
should become observable, provided the disorder effects do not
intervene. The crossover from high to low-density behavior is discussed
in more detail in the next section where we give arguments that the
strong spin-charge separation sets in already at rather modest $r_s$.


\section{Validity domain of the obtained formulas and numerical
estimates}
\label{sec:Domain}
\subsection{Intermediate densities}

Returning to the case of the unscreened Coulomb interaction, an
important practical question is the relevance of the obtained formulas
for a more common experimental situation of $r_s \sim 1$. Although
Eq.~(\ref{eqn:J_from_g_ultrathin}) reduces this question to an easier
problem of calculating the PCF of spinless fermions, at such $r_s$ it
still has to be done numerically. Deferring this effort for a possible
future work, we attempted to see whether one can estimate $J$ at $r_s
\sim 1$ by a simple interpolation between the high-density limit
$r_{s}\ll 1$, corresponding to the CTG,\cite{Fogler-05a} and the
low-density WC limit, $r_s \gg 1$, studied in this work.

\begin{figure}
\begin{center}
\includegraphics[height=1.75in]{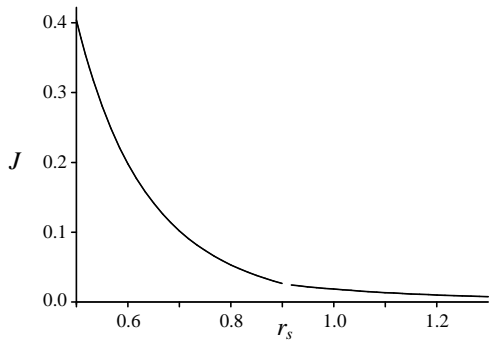}
\end{center}
\caption{Exchange constant in units of $e^2/\epsilon a_{B}$ as a function
of $r_s$ for $a_{B}/R = 100$. The left curve is computed according to
Eq.~(\ref{eqn:J_CTG}), the right one is per Eq.~(\ref{eqn:J}).}
\label{fig:J}
\end{figure}

Using Eq.~(\ref{eqn:J_CTG_from_t}) with $\tilde{g}_0(q)$ computed to the
linear order in $r_s$ (per Ref.~\onlinecite{Fogler-05a}) we obtained
for the CTG
%
\begin{equation}
J = \frac{e^2}{\epsilon a_B} \frac{\pi^2}{24 r_s^3 \mathcal{L}}
\left[
1 - \frac{2 (3 + \pi^2)}{3 \pi^2} r_s
\right] + O\left(\frac{1}{\mathcal{L}^2}\right),
\label{eqn:J_CTG}
\end{equation}
which is valid for ${1}/{\mathcal{L}} \ll r_s \ll 1$. We plotted this
dependence in Fig.~\ref{fig:J} together with the prediction of
Eq.~(\ref{eqn:J}) for the infinite-wire WC. As one can see, a smooth
interpolation between the two is entirely possible. In fact, the curves
match almost seamlessly around $r_s \sim 0.9$ where a tiny gap has been
left on purpose to separate them from each other. However, we must warn
the reader that at such $r_s$ the high-density approximation of
Eq.~(\ref{eqn:J_CTG}) certainly needs higher-order corrections (taken
literally, it would give negative $J$ at $r_s > 1.15$). Similarly,
Eq.~(\ref{eqn:J}) is likely to have its own non-negligible corrections
at such $r_s$. Thus, one should regard Fig.~\ref{fig:J} as the
indication that the crossover from the CTG to the WC occurs at $r_s =
1$--$2$. To put it another way, we expect that our Eq.~(\ref{eqn:J})
becomes quantitatively accurate at $r_s \gtrsim 2$. This conclusion
seems to be in agreement with the numerical result of
Ref.~\onlinecite{Egger-99} that the 1D WC is well-formed already at $r_s
> 4$. To get an idea of the size of $J$, we can use parameters $r_s =
4$, $a_B = 1.5$~nm, and $\epsilon = 1$, which are relevant for the
experiment of \textcite{Jarillo-Herrero-04} For this set of numbers
Eq.~(\ref{eqn:J}) yields
\begin{equation}
J \approx 1\ \text{K}.
\end{equation}
Unfortunately, the lowest temperature in that experiment was $0.3$~K, so
the exchange correlations may have been strongly affected. We leave the
investigation of the finite-temperature effects for a future study. We
also hope that lower temperatures can be achieved in the next round of
experiments, so that we would be in a better position to check our
predictions.

%

\subsection{Thicker wires}

Another question that we wish to briefly address is the behavior of the
exchange constant in a wire of radius $R \gtrsim a_B$. This parametric
regime is more common than that of ultrathin wires, $R \ll a_B$, and so
it needs to be investigated.

In regards to the WC limit, the correction to parameter $\eta$ of
Eq.~(\ref{eqn:J}) due to a finite $R$ has been studied by
\textcite{Klironomos-05} These authors pointed out that in the wires
with radius $R\gtrsim a_{B}$ the problem of spin exchange cannot be
treated as a purely 1D one from the outset. They found that the
tunneling trajectories of the primary electron pair avoid the head-on
collision by deviating off the wire center in the transverse direction
by a certain amount $r_0$ each. For the parabolic confinement potential
$U_{\text{con}}(r_\perp) = (\hbar^2 / 2 m R^2) (r_\perp / R)^2$ it is
easy to find the estimate $r_0 \sim (R^4 / a_B)^{1/3}$ by balancing the
Coulomb repulsion and the lateral confinement forces at the point of the
closest approach. As long as $r_0 \ll a$, it remains possible to modify
the interaction potential $U(x)$ to take this effect into account. [At
the simplest level, it is sufficient to replace $R$ by $2 r_0$ in
Eq.~(\ref{eqn:U}).] Thereafter, one can still treat the problem as
1D.\footnote{
Significant changes in the calculation would be required only in very
thick wires with radius $R > 11\,a_{B}$.\cite{Klironomos-05} However, in
such wires the 1D WC limit can be achieved only at $r_s \gtrsim 20$
where the exchange coupling $J$ would be unmeasurably small, see
Eq.~(\ref{eqn:J}). At smaller $r_s$, other physics is likely to come
into play, e.g., the 1D WC may undergo a transition into a multi-row
WC\cite{Piacente-04, Klironomos-xxx} or into a liquid with multiple
subband occupation.
}

In thick wires the relationship between the PCF and the exchange constant
would differ from Eq.~(\ref{eqn:J_from_g_ultrathin}). Of
primary importance here is the small-distance behavior of $U(x)$. When
$r_0 \gg a_B$, the interaction potential becomes so smooth near $x = 0$
that it can be approximated by a constant $U(0) \sim e^2 / 2 \epsilon
r_0$. Following the method outlined in Appendix~\ref{sec:U_sigma}, one
arrives at the relation
\begin{equation}
J = \frac{e^2 a_B}{2 \epsilon} a_0 g_0^{\prime \prime }(x),
\quad a_0 = \frac{\hbar}{\sqrt{m U(0)}} \sim (R^2 a_B)^{1/3}.
\label{eqn:J_smooth}
\end{equation}
We see that the factor $\mathcal{L}$ in
Eq.~(\ref{eqn:J_from_g_ultrathin}) gets replaced with
$a_B / a_0 \ll 1$.\footnote{
It is interesting that in the opposite limit of a short-range $U(x)$,
e.g., $U(x) = c \delta(x)$, the relation between $J$ and $g_0$ is
formally similar, with $a_0 = 2 \hbar^2 / m c$, see
Eqs.~(\ref{eqn:J_CTG_from_t}) and (\ref{eqn:t_q}).
}

In the case of intermediate-width wires, $R \sim a_B$, one can use $a_0
= a_B$ for order-of-magnitude estimates; however, the exact value of the
coefficient $a_0$ and therefore $\kappa$ in Eq.~(\ref{eqn:J}) would
depend on the details of the confinement potential
$U_{\text{con}}(\textbf{r}_\perp)$.


\section{Conclusions}
\label{section:Conclusions}

In this paper we have developed a method for a controlled calculation of
the exchange constant $J$ in correlated 1D electron systems to the
leading order in the large parameter $r_s$. This method is able to
reproduce a number of exact results for integrable models, which is a
strong indicator of its validity. In comparison to a prior uncontrolled
calculation\cite{Matveev-04} of the same quantity for the Coulomb case,
our method brings a small correction of about $0.7\%$ to the coefficient
in the exponential term; however, the correction to the pre-exponential
factor is large, about 50\%.

Our another main result --- the relation between $J$ and the
pair-correlation function of spinless electrons --- is a promising
ground for an attack on the regime of intermediate concentrations, $r_s
\sim 1$, which has so far been difficult for a controlled analytical
study. We demonstrated that a very smooth interpolation between the high
and low-density asymptotic is possible based already on the results in
hand (Fig.~\ref{fig:J}).

Finally, we have applied our method to a number of realistic interaction
potentials corresponding to different geometries of experimental setup.
Apart from nearly impenetrable Coulomb interaction in ultrathin quantum
wires and quantum rings, which has been the primary focus of this work,
we have studied penetrable smooth interactions that take place in the
wires of intermediate width and a screened $1/x^2$ interaction that is
realized, e.g., when the wire is located near a metallic half-plane.

These findings have direct implications for the energy spectroscopy of
spin-splitting in nanoscale quantum rings\cite{Lorke-00, Fuhrer-04,
Bayer-03} and 1D quantum dots.\cite{Field-90, Jarillo-Herrero-04,
Auslaender-05} Some estimates were given in
Secs.~\ref{sec:integrable_model} and \ref{sec:Domain} above. They should
be especially accurate in the ultrathin-wire limit\cite{Fogler-05a}
$\mathcal{L} = \ln (a_B / R) \gg 1$ that can be achieved in
carbon-nanotube field-effect transistors with high-$k$
dielectrics.\cite{Javey-02}

In long 1D wires, the exchange coupling $J$ determines the velocity
of the elementary spin excitations
\begin{equation}
v_{\sigma } = \frac{\pi}{2} \frac{J a}{\hbar},
\end{equation}
which can nowadays be measured by tunneling,\cite{Auslaender-05}
angle-resolved photoemission,\cite{Claessen-02} or deduced from the
enhancement of the spin susceptibility and electron specific
heat.\cite{Fogler-05a} Our result for $v_{\sigma }$ reads
(cf.~Table~\ref{tbl:Results})
\begin{equation}
{v_{\sigma }}/{v_{F}}=17.8\,{r_{s}^{3/4}}e^{-\eta \sqrt{2r_{s}}}/
{\mathcal{L}},\quad \eta =2.7978(2),  \label{eqn:v_sigma_Coulomb}
\end{equation}
where $v_F = (\pi /2)(\hbar / m a)$ is the Fermi velocity.


\begin{acknowledgments}

The support from the A.~P. Sloan Foundation and C.~\&~W. Hellman Fund is
gratefully acknowledged. We thank A.~D. Klironomos, R.~R. Ramazashvili,
and K.~A. Matveev for discussions.

\end{acknowledgments}

\appendix

\section{Spin-dependent part of the effective potential}
\label{sec:U_sigma}

In this Appendix we derive Eqs.~(\ref{eqn:U_sigma}),
(\ref{eqn:U_sigma_CSM}), and (\ref{eqn:J_smooth}). Let $x \equiv
x_1 - x_0$ be the relative distance between the two electrons for which
we want to compute the exchange coupling. Consider the orbital part of
the many-body wavefunction $\Phi$,which is a function of $x$ and $N - 1$
other coordinates. In principle, $\Phi$ is different for different
many-body spin states. Nonetheless, under the condition $J \ll E_F$
where only nearest-neighbor exchanges are important, at small $x$
function $\Phi$ depends primarily on the total spin of the given pair.
The triplet-state wavefunction $\Phi_t(x)$ is odd and therefore
$\Phi_t(0) = 0$. In the singlet state we have $\Phi_s(x)$ which is even
and finite at $x = 0$ (cf.~Sec.~\ref{sec:Three}). However, because of
the strong Coulomb repulsion $\Phi_s(0)$ is very small. At $x\ll a$ the
two electrons of interest are under a large Coulomb barrier and it is
legitimate to assume that $\Phi_t$ and $\Phi_s$ are functions of only
$x$, the fastest variable in the problem. Our goal is to choose
$U_{\sigma}(x)$ that reproduces the difference between $\Phi_{s}$ and
$\Phi_{t}$ at the level of the first-order perturbation theory. This is
achieved if there exists a window $a_{c}\ll x\ll a$, where the following
equation is satisfied:
\begin{equation}
\Phi_{t}(x) - \Phi_{s}(x) = \int\limits_{0}^{\infty }dx^{\prime}
G(x,x^{\prime})U_{\sigma }(x^{\prime })\Phi_{t}(x^{\prime }).
\label{eqn:U_sigma_eqn}
\end{equation}
Here the Green's function $G$ is the solution of
\begin{equation}
\left[ -\frac{\hbar ^{2}}{2\mu }\partial _{x}^{2}+U(x)-E\right]
G(x,x^{\prime })=-\delta (x-x^{\prime }),
\label{eqn:G_equation}
\end{equation}
where $E\sim e^{2}/\epsilon a$ is the total energy. At the end of the
calculation one can verify that the final result
[Eq.~(\ref{eqn:U_sigma})] holds independently of the precise value of
$E$ up to terms $O(1/\mathcal{L})$.

Our next step is to use the usual representation of $G$,
\begin{equation}
G(x,x^{\prime })=-\frac{2\mu }{\hbar ^{2}Q}\left\{
\begin{array}{ll}
u(x)v(x^{\prime }), & x>x^{\prime }, \\
v(x)u(x^{\prime }), & x<x^{\prime },
\end{array}
\right.
\end{equation}
in terms of two linearly-independent solutions of the (homogeneous)
Schr\"odinger equation~(\ref{eqn:G_equation}). The solution $u(x)$
initially decays exponentially with $x$ and then turns into a wave
propagating towards $x=+\infty $. The other solution, $v(x)$ has a node
at $x = 0$, exhibits an exponential rise and finally becomes a standing
wave. The quantity $Q = u v^{\prime} - u^{\prime } v$ is their Wronskian.

Wavefunctions $\Phi_{s}$ and $\Phi_{t}$ are the linear combinations of
$u$ and $v$ that are even and odd in $x$, respectively. For $x$ at which
Eq.~(\ref{eqn:U_sigma_eqn}) is valid, $\Phi_{t}\simeq \Phi_{s}\simeq
\text{const}\times v(x)$. Using this property and some trivial algebra,
we get
\begin{equation}
\int\limits_{0}^{\infty} dx U_\sigma(x)v^2(x)=\frac{\hbar^2}{2\mu }%
\frac{Q^{2}}{u(+0)u^{\prime }(+0)}.
\label{eqn:U_sigma_II}
\end{equation}
For the Coulomb case [Eq.~(\ref{eqn:U})] $u$ and $v$ can be expressed in
terms of Whittaker functions\cite{Gradshteyn-Ryzhik} $W_\alpha(z)$ and
$M_\alpha(z)$ (see an earlier remark in Sec.~\ref{sec:Three} and
Ref.~\onlinecite{Fogler-05a}):
\label{eqn:u,v_Coulomb}
\begin{equation}
\begin{split}
u(x)& =W_{-i\nu, 1/2}[i(x-R)/b], \\
v(x)& =M_{-i\nu, 1/2}[i(x-R)/b]  \\
& -M_{-i\nu, 1/2}(-iR / b)u(x)/u(0),
\end{split}
\end{equation}
where $b = \hbar / (8\mu E)^{1/2}$ and $\nu = b / a_{B}\sim
\sqrt{r_{s}}\gg 1$. Using the relations\cite{Gradshteyn-Ryzhik}
$W_{-i\nu ,1/2}(0)=1/\Gamma (1+i\nu )$ and $M_{-i\nu,1/2}(iz) \simeq iz$
at $z\ll 1$, we find $Q=i/\Gamma (1-i\nu) b$. Substituting these
results into Eq.~(\ref{eqn:U_sigma_II}) and dropping terms that are
small in the parameter $R/a_{B}\ll 1$, we recover
Eq.~(\ref{eqn:U_sigma}).

Similarly, for the CSM model [Eq.~(\ref{eqn:U_CSM})] $u$ and $v$ are
given by
\begin{equation}
\begin{split}
u\left( x\right) & =\sqrt{x}H_{\lambda -1/2}^{(1)}(qx), \\
v\left( x\right) & =\sqrt{x}J_{\lambda -1/2}(qx),
\end{split}
\end{equation}
where $q=\sqrt{m E}/\hbar$, and $H_\nu^{(1)}$, $J_\nu$ are the Hankel
and the Bessel functions of the first kind,\cite{Gradshteyn-Ryzhik}
respectively. Their Wronskian is given by $W\{u,v\}=1/i\pi (\lambda -
1/2)$, and their known asymptotic behavior leads in this case
to the formulas
\begin{equation}
\begin{split}
u(R) &\simeq \Gamma (\lambda -1/2) R^{1-\lambda }/i\pi,
\\
u^{\prime}(R) &\simeq (1-\lambda) u(R) / R,
\\
v(R) & \simeq R^{\lambda }/\Gamma (\lambda +1/2),
\end{split}
\end{equation}
so that the constraint on $U_\sigma(x)$ is expressed by
Eq.~(\ref{eqn:U_sigma_CSM}).

Finally, in thick 1D wires, where $U(x) \simeq \text{const}$ at small
$x$ (Sec.~\ref{sec:Domain}), the solutions of Eq.~(\ref{eqn:G_equation})
are $u(x) = \exp(-x / a_0)$ and $v(x)=\sinh (x / a_0)$, where $a_0 =
\hbar / \sqrt{m U(0)}$. These formulas leads to
Eq.~(\ref{eqn:J_smooth}).

\section{Physical realization of the Calogero-Sutherland-Moser model}
\label{sec:CSM_law}

\begin{figure}
\begin{center}
\includegraphics[height=1.75in]{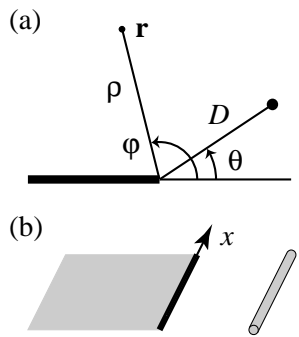}
\end{center}
\caption{The geometry of the electrostatic problem that
gives rise to the CSM interaction law. (a) The cross-sectional view
in a plane normal to the $x$-axis. The thick line symbolizes the metal
half-plane, the large dot stands for the 1D wire. The drawing also
indicates how the polar coordinates $\rho$, $\varphi$ of an arbitrary point
$\textbf{r}$ are defined. (b) A three-dimensional view of the system.}
\label{fig:CSM}
\end{figure}

The CSM interaction law can be realized experimentally if the 1D wire is
positioned nearby a metallic half-plane, parallel to its edge, see
Fig.~\ref{fig:CSM}. In this configuration the interaction potential has
the Coulomb form [Eq.~(\ref{eqn:U})] only at small $x$. Below we show
that at larger distances, $|x| \gg D$, the interaction changes to the
CSM law $U(x) \simeq A / x^2$, due to the screening effect of the metal.
For simplicity, let us assume that the metallic half-plane is an ideal
conductor and that $x \gg R$. In this case $U(x) = V(x, D, \theta)$,
where $V$ is the solution of the following electrostatic problem:
\begin{align}
\nabla^2 V(\textbf{r}) &= \left(\frac{\partial^2}{\partial x^2} +
 \frac{1}{\rho} \frac{\partial}{\partial \rho} \rho
 \frac{\partial}{\partial \rho}
 + \frac{1}{\rho^2} \frac{\partial^2}{\partial \varphi^2}\right)
 V(x, \rho, \varphi)
\nonumber\\
 &= -\frac{4 \pi e^2}{\epsilon D} \delta(x) \delta(\rho - D)
    \delta(\varphi - \theta),
\label{eqn:Laplace_V}\\
V(\textbf{r}) &= 0,\quad \varphi = \pm \pi.
\label{eqn:bc_V}
\end{align}
Using standard methods (cf.~Ref.~\onlinecite{Fogler-04} and references
therein) one can express $V$ in terms of elliptic integrals. However,
for a general $\theta$, such expressions are not very illuminating. An
exception is the case $\theta = 0$, where $V$ takes the form
\begin{equation}
V(\textbf{r}) = \frac{2}{\pi \epsilon} \frac{e^2}{|\textbf{r} - \textbf{D}|}
\arctan \frac{\sqrt{2 D \rho \left( 1 + \cos\varphi\right) }}
{|\textbf{r} - \textbf{D}|},
\label{eqn:V_00}
\end{equation}
where $\textbf{D} = (0, 0, D)$. This formula can be verified by the direct
substitution into Eqs.~(\ref{eqn:Laplace_V}) and (\ref{eqn:bc_V}). For the
intra-wire interaction law we obtain
\begin{equation}
U(x) = \frac{2}{\pi \epsilon} \frac{e^2}{|x|} \arctan \frac{2D}{|x|},
\label{eqn:U_00}
\end{equation}
so that $A = 4 D e^2 /\pi \epsilon$ in this case. We can also show that
for an arbitrary $-\pi < \theta < \pi$ the coefficient $A$ is given by
\begin{equation}
A = \frac{2 D}{\pi} \frac{e^2}{\epsilon} \left( 1 + \cos\theta\right) .
\label{eqn:A}
\end{equation}
First we note that the Fourier transform of $U$ must have the form
$\tilde{U}(q) \simeq \tilde{U}(0) - \pi A |q|$ at small $q$. On the
other hand, $\tilde{U}$ is given by the generalized Fourier series
%
\[
\tilde{U}(q) = \frac{2 e^2}{\epsilon} \sum\limits_{n = 1}^\infty
\left[1 - (-1)^n \cos n \theta\right] I_{{n}/{2}} (q D)
K_{{n}/{2}}(q D),
\]
%
where $I_\nu$ and $K_\nu$ are the modified Bessel functions of the first
and the second kind, respectively.\cite{Gradshteyn-Ryzhik} The
non-analytic $|q|$-contribution comes only from the $n = 1$ term in the
series. Using the formulas\cite{Gradshteyn-Ryzhik} $I_{1/2}(z) = (2 /
\pi z)^{1/2} \sinh z$ and $K_{1/2}(z) = (\pi / 2 z)^{1/2} \exp(-z)$, we
arrive at Eq.~(\ref{eqn:A}).

%

The geometry of Fig.~\ref{fig:CSM}(b) has been realized in the experiment
of \textcite{Auslaender-05}, where the role of the
metallic half-plane was played by a two-dimensional electron gas. In
contrast to our idealized electrostatic model, the two-dimensional metal
has a non-zero Thomas-Fermi screening radius $r_{\text{TF}}$. We expect that in
this situation the only change in Eq.~(\ref{eqn:A}) is the replacement of
$D$ by $D + r_{\text{TF}}$. This yields the following estimate for
the parameter $\lambda$ in Eq.~(\ref{eqn:U_CSM}):
\begin{equation}
\lambda = \frac12 + \sqrt{\frac14 + \frac{2}{\pi}
          \frac{D + r_{\text{TF}}}{a_B} (1 + \cos\theta)}.
\label{eqn:lambda}
\end{equation}
In the experimental setup of Ref.~\onlinecite{Auslaender-05} $D \sim a_B
\sim r_{\text{TF}}$, so that $\lambda = 1.5$--$2$.


\end{document}